# Reduced-Complexity Column-Layered Decoding and Implementation for LDPC Codes

Zhiqiang Cui[1], Zhongfeng Wang[2], *Senior Member, IEEE*,  and Xinmiao Zhang[3]

[1]Qualcomm Inc., San Diego, CA 92121, USA

[2]Broadcom Corp., Irvine, CA 92617, USA

[3]Case Western Reserve University, Cleveland, OH 44106, USA

Abstract: Layered decoding is well appreciated in Low-Density Parity-Check (LDPC) decoder implementation since it can achieve effectively high decoding throughput with low computation complexity.  This work, for the first time, addresses low complexity column-layered decoding schemes and VLSI architectures for multi-Gb/s applications. At first, the Min-Sum algorithm is incorporated into the column-layered decoding. Then algorithmic transformations and judicious approximations are explored to minimize the overall computation complexity. Compared to the original column-layered decoding, the new approach can reduce the computation complexity in check node processing for high-rate LDPC codes by up to 90% while maintaining the fast convergence speed of layered decoding. Furthermore, a relaxed pipelining scheme is presented to enable very high clock speed for VLSI implementation. Equipped with these new techniques, an efficient decoder architecture for quasi-cyclic LDPC codes is developed and implemented with $0.13um$ CMOS technology. It is shown that a decoding throughput of nearly 4 Gb/s at maximum of 10 iterations can be achieved for a (4096, 3584) LDPC code. Hence, this work has facilitated practical applications of column-layered decoding and particularly made it very attractive in high-speed, high-rate LDPC decoder implementation.

## 1 INTRODUCTION

Conventionally, LDPC codes are decoded using the Sum-Product algorithm (SPA) [1] or the modified Min-Sum algorithm (MSA) [2]. In general, the SPA has the best decoding performance. The MSA is an approximation of the SPA aimed to reduce computation complexity. It is widely employed in LDPC decoder design [3][4][5][6]. Both algorithms are based on two-phase message-passing (TPMP) scheme. In the literature, variations of the SPA and MSA decoding approaches have been investigated to reduce the interconnect complexity in VLSI implementation [15][16]. Recently, layered LDPC decoding schemes [7][8][9][10] have attracted much attention in both academy and industry because they can effectively speed up the convergence of LDPC decoding and thus reduce the required maximum number of decoding iterations. Presently two kinds of layered decoding approaches, i.e., row-layered decoding [7][8][9] and column-layered decoding [9][10] have been proposed. In row-layered decoding, the parity check matrix of the LDPC code is partitioned into multiple row layers. The message updating is performed row layer by row layer. The column-layered decoding employs the similar idea except that the parity check matrix is partitioned into multiple column layers. The message computation is performed column layer by column layer.

The implementation of LDPC decoder with row-layered decoding has been widely studied [11][12][13][14]. The SPA-based column-layered decoding approach [10] was proposed in 2005. In the same year, Radosavljevic [9], *et al,* proposed a simplification of the SPA-based column-layered decoding approach. It has been reported that the column-layered decoding has a similar convergence speed as row-layered decoding. However, the column-layered decoding algorithm has attracted much less attention due to its inherent high computation complexity. In this work, we investigate and explore the benefits of the column-layered decoding approach. We first incorporate the Min-Sum algorithm into the column-layered message passing scheme because it has much lower complexity than the SPA. In addition, by deeply investigating the message updating process in the Min-Sum based column-layered decoding, we develop a simplified column-layered decoding scheme, which maximally eliminates the redundant computations in the original scheme and significantly reduces the computation complexity further with judicious algorithmic approximation. It is shown that up to 90% of computations in check node processing can be saved for high rate LDPC codes.

For high-speed VLSI implementation, the proposed design has significant advantages over the conventional row-layered decoding. First, the column-layered decoding inherently has shorter critical path. For a row-layered decoder, the messages associated to multiple sub-blocks in a row layer can be processed in one cycle to increase throughput. However, it will increase the complexity of check node unit (CNU) and require serial concatenation of multiple comparison and selection stages in VLSI implementation. It has been reported that significant hardware overhead is required to optimize the corresponding circuitry for high clock speed [12]. In the column layered decoding, the major implementation complexity is associated with variable node unit (VNU), particularly when the messages corresponding to multiple sub-blocks in a column layer are processed in parallel. Because only addition operations are performed in a VNU, it is very convenient to employ arithmetic optimization to minimize the critical path. Second, in the proposed design, the overall pipeline latency in decoding a code block is equal to the number of pipeline stages while in row-layered decoding, the same amount of pipeline latency is introduced for the message updating of every layer. Moreover, the intrinsic message loading latency is minimized in the column-layered decoding because the decoding can start as soon as the intrinsic messages corresponding to one block column are available. In summary, the proposed design is well suited for very high decoding throughput and low power LDPC decoder implementation.

The column-layered decoding scheme (also known as the shuffled iterative decoding) for LDPC codes was proposed in [10]. The decoding scheme is based on the SPA algorithm. Similar to row-layered decoding [7][8][9], the maximum number of decoding iterations can be significantly reduced for the same decoding performance. Since the Min-Sum algorithm has much lower implementation complexity than the SPA algorithm, it is widely utilized in hardware implementation. In this paper, the Min-Sum algorithm is incorporated into the column-layered decoding for the first time. Then, algorithmic transformations and intelligent approximations are explored to significantly reduce the computation complexity and memory requirement.

## 2.1 Min-Sum Algorithm Based Column-layered Decoding

Let $C$ be a binary (N, K) LDPC code specified by a parity-check matrix $\boldsymbol{H}$ with $M$ rows and $N$ columns. Each row of the parity check matrix is associated with a check node, and each column is associated with a variable node. Let $N(c) = \{v : H_{cv} = 1\}$ denote the set of variable nodes that participate in check node $c$, and $M(v) = \{c : H_{cv} = 1\}$ denote the set of check nodes associated to variable node $v$. Let $I_v$ denote the intrinsic message for variable node v and $R_{cv}$ represent the check-to-variable message conveyed from check node $c$ to variable node $v$, and $L_{cv}$ represent the variable-to-check message conveyed from variable node $v$ to check node $c$. Assume that the $N$ bits of a codeword are divided into $G$ groups of the same size, $N_0, N_1, \cdots N_{G-1}$. Accordingly, the parity-check matrix is divided into $G$ block columns. The proposed column-layered decoding based on the Min-Sum algorithm is described with the pseudo code below.

### Min-Sum-based column-layered decoding algorithm

---

**Initialization:**

$L_{cv} = I_v$ for $v$=0, 1, ..., N-1, c=0, 1, ..., M-1;

**Iterative decoding:**

For *iter* = 1, 2, …, maximum iteration number

{

    For *g*=0, 1, …, *G*-1

    {

        *Horizontal Step*: For each check node *c* that is connected to variable node $v \in N_g$ , computes

$$R_{cv} = \prod_{n \in N(c) \setminus v} \mathrm{sgn}(L_{cn}) \times \min_{n \in N(c) \setminus v} |L_{cn}|. \quad (1)$$

        *Vertical Step*: For each variable node $v \in N_g$ , updates $L_{cv}$ and $L_v$ as follows:

$$L_{cv} = I_v + \alpha \times \sum_{m \in M(v) \setminus c} R_{mv} , \quad (2)$$

$$L_v = I_v + \alpha \times \sum_{m \in M(v)} R_{mv} . \quad (3)$$

    }

    **Hard decision and termination:**

    Make hard decision by using the sign of $L_v$ ; Terminate the decoding if a valid codeword is found.

}

---

The optimum value of the scaling factor $\alpha$ is around 0.8[Wrong!]0. For the convenience of VLSI implementation, it is set as 0.75 in this work.

Assume all variable nodes are divided into 4 groups (*i.e.*, *G* = 4). The computation flow of the column-layered decoding for one iteration is illustrated in Figure 1. (a). The shaded sub-matrices indicate coverage of computation in decoding each layer, where the computation in (1) must be carried out for all block columns except for the current updating block column. Hence, the computation complexity of the original column-layered decoding scheme per iteration is many times more than that of the conventional TPMP scheme whether the MSA or SPA is used.

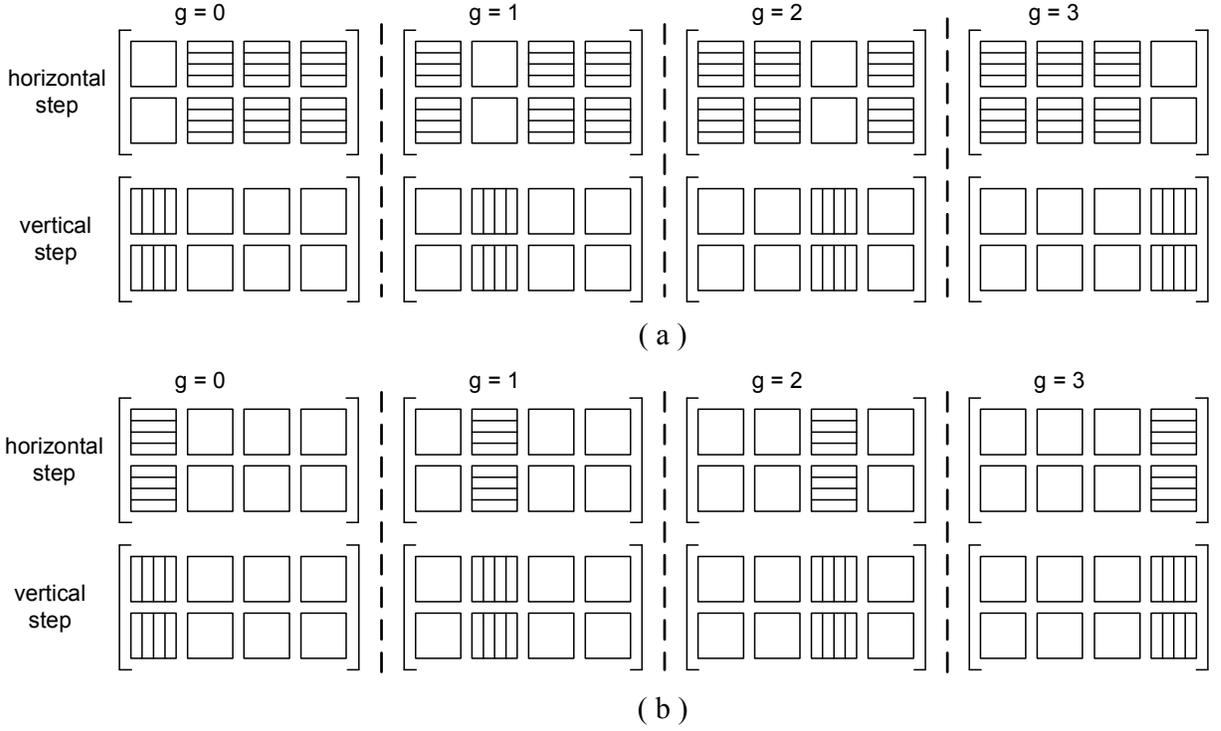

Figure 1.　　The computation flow in an iteration of column-layered decoding,
(a) original algorithm, (b) after algorithm reforumlation.

### 2.2 Low Complexity Decoding Scheme

#### 2.2.1 Algorithm Reformulation

With a close study of the computation flow of the column-layered decoding, it can be observed that a significant amount of redundant computation is performed in the original column-layered decoding algorithm. To improve the computation efficiency, the computation in (1) for consecutive column layers can be incrementally performed. In LDPC decoding, each variable node sends a variable-to-check message to every neighboring check node. Assume that a check node $c$ has $d_c$ variable node neighbors, and hence receives $d_c$ soft messages from its neighboring variable nodes. To clarify the main procedure of the reformulated column-layered decoding, we assume that each check node has only one variable node neighbor in each block column of the parity check matrix. It should be noted that this constraint is not indispensable for column-layered decoding in general. Let $\mathbf{m}_c^{(g)} = [m_1 m_2 ... m_{d_c}]$ be a sorted vector of the magnitudes of the soft messages received by check node $c$ in ascending order. The superscript $g$ indicates the soft message was generated when the $g^{th}$ block column is processed. Similarly, let $S_c^{(g)}$ be $\prod_{n \in N(c)} sign(L_{cn})$ when the decoding for the layer $g$ is completed. To reduce the computation complexity of Min-Sum based column-layered decoding, $\mathbf{m}_c^{(g)}$ can be computed from $\mathbf{m}_c^{(g-1)}$ in three steps.

1. For each variable node $v$ in $N_g$, remove the old $|L_{cv}|$ from $\mathbf{m}_c^{(g-1)}$ to obtain a temporary sorted vector $\tilde{\mathbf{m}}_c^{(g)}$. In addition, remove the old $sign(L_{cv})$ from $S_c^{(g-1)}$ and obtain a temporary sign-product $\tilde{S}_c^{(g)}$. Since the smallest value, $m_1$, in $\tilde{\mathbf{m}}_c^{(g)}$ is $\min_{n \in N(c) \backslash v}(|L_{cn}|)$ and $\tilde{S}_c^{(g)}$ is $\prod_{n \in N(c) \backslash v} sign(L_{cn})$, the value of $R_{cv}$ can be computed as $\tilde{S}_c^{(g-1)} \times m_1$. Send the $R_{cv}$ message to the variable node $v$.

2. Perform variable-to-check message computations for all variable nodes belonging to $N_g$. The new $L_{cv}$ messages are sent back to corresponding check nodes.

3. For each check node, insert the updated $|L_{cv}|$ into $\tilde{\mathbf{m}}_c^{(g)}$ in a sorted order to obtain $\mathbf{m}_c^{(g)}$.

The reformulated column-layered decoding procedure is summarized as follows:

**The Reformulated Column-layered Decoding**

---

**Initialization:**

Let $L_{cv} = I_v$ for all variable nodes. For each check node, sort the magnitudes of the $L_{cv}$ messages from its neighboring variable nodes. Compute the sign product for each check node $S_c = \prod_{n \in N(c)} sign(L_{cn})$.

**Iterative decoding:**

For $iter$ = 1, 2, ..., maximum iteration number

{

   For g=0, 1, ..., *G-1*

   {

       ***Horizontal Step-A***: for each check node $c$ that connects to variable node $v \in N^{(g)}$,

       compute $\tilde{\mathbf{m}}_c^{(g)}$ by removing the old $|L_{cv}^{(g)}|$ from $\mathbf{m}_c^{(g-1)}$,                      (4)

       and $R_{cv}^{(g)} = \tilde{S}_c^{(g)} \times m_1$, $where$ $\tilde{S}_c^{(g)} = S_c^{(g-1)} \times old\, sign(L_{cv}^{(g)})$.               (5)

       ***Vertical Step***: For each variable node $v \in N_g$, compute $L_{cv}^{(g)}$ and $L_v^{(g)}$ using (2) and (3).

***Horizontal Step-B***: for each check node $c$ that connects to variable node $v \in N^{(g)}$,

compute $\mathbf{m}_c^{(g)}$ using $\tilde{\mathbf{m}}_c^{(g)}$ and $\left| L_{cv}^{(g)} \right|$, $\hspace{3cm}$ (6)

and $S_c^{(g)} = \tilde{S}_c^{(g)} \times sign(L_{cv}^{(g)})$. $\hspace{3cm}$ (7)

}

**Hard decision and termination:**

Make hard decision by using the sign of $L_v$; Terminate decoding if a valid codeword is found or the maximum decoding iteration is reached.

}

In the above algorithm, $\tilde{\mathbf{m}}_c^{(0)}$ and $\tilde{S}_c^{(0)}$ in a decoding iteration are computed from $\mathbf{m}_c^{(G-1)}$ and $S_c^{(G-1)}$, respectively, which are obtained in the previous iteration. The new computation flow for one full iteration is depicted in Figure 1. (b).

### 2.2.2 *Simplification of Min-Sum Based Column-layered Decoding*

The algorithm reformulation presented above removes a large amount of redundant computation in the original column-layered decoding scheme, and thus significantly reduces the overall computation complexity. However, because every vector $\mathbf{m}_c^{(g)}$ contains $d_c$ values, the reformulated algorithm still requires a considerable amount of memory to store check-to-variable messages. For row $c$, only the two smallest values in the sorted vector $\mathbf{m}_c^{(g)}$ are directly involved in the message updating Step-A from layer $g$-$1$ to layer $g$. For example, if the smallest value in $\mathbf{m}_c^{(g-1)}$ is from layer $g$ in the previous iteration, it is removed from the vector $\mathbf{m}_c^{(g-1)}$ in horizontal Step-A. Then the second smallest value is used as the magnitude of variable-to-check message. In horizontal Step-B, $\mathbf{m}_c^{(g)}$ is computed with the new value, $| L_{cv} |$, from variable node, and the pre-sorted vector $\tilde{\mathbf{m}}_c^{(g)}$ from horizontal Step-A. The new value, $| L_{cv} |$, could take any index in $\mathbf{m}_c^{(g)}$. If it takes a very small index, it has more chance to be used in Step-A of further computation. Otherwise, it has much less chance to be one of the two smallest values in the remaining computation. In another word, though $\mathbf{m}_c^{(g)}$ contains $d_c$ values, most of the values in the end of the sorted vector are less likely to be used as reliability information for message updating. Thus, it is reasonable to reduce the length of the vector $\mathbf{m}_c^{(g)}$ to further reduce the implementation complexity. Our simulations show that if the lengths of the vectors $\mathbf{m}_c^{(g)}$ and $\tilde{\mathbf{m}}_c^{(g)}$ are set to be 3, the decoding convergence speed and performance have almost no degradation compared to the standard Min-Sum based column-layered decoding. In such case, we name the scheme as *three-min column-layered decoding*. Similarly, the lengths of the vectors $\mathbf{m}_c^{(g)}$ and $\tilde{\mathbf{m}}_c^{(g)}$ can be set as 2 or even 1. Correspondingly, more penalties in convergence speed and performance are expected.

In this approximation, $\tilde{\mathbf{m}}_c^{(g)}$ contains three values in most cases. It may contain two valid values if one value is removed from the vector $\mathbf{m}_c^{(g-1)}$ in horizontal Step-A. In the computation of horizontal Step-B, at most three comparison operations are required to sort the new value $|L_{cv}|$ from variable node and the sorted values in $\tilde{\mathbf{m}}_c^{(g)}$. If $\tilde{\mathbf{m}}_c^{(g)}$ contains three valid values, the third comparison is needed to determine whether the $|L_{cv}|$ is the third smallest value or it should be discarded. Table I shows the average number of $\mathbf{m}_c^{(g)}$ updating events in one iteration for the decoding of a (4096, 3584) (4, 32) QC-LDPC code at the SNR of 4.1dB. In one iteration, the total number of $\mathbf{m}_c^{(g)}$ updating events for a check node is 32. A message updating event can be categorized into one of three types. Type I, a value is removed from $\mathbf{m}_c^{(g-1)}$ and then a new value is inserted into $\tilde{\mathbf{m}}_c^{(g)}$. Type II, no value is removed from $\mathbf{m}_c^{(g-1)}$ but $\mathbf{m}_c^{(g)}$ is updated. Type III, no value is removed from $\mathbf{m}_c^{(g-1)}$ and $|L_{cv}|$ is discarded. It can be seen from Table I that if no value is removed from $\mathbf{m}_c^{(g-1)}$, $|L_{cv}|$ is much more likely to be discarded than being the third smallest value in $\mathbf{m}_c^{(g)}$. To further reduce the decoding complexity, the third comparison in horizontal Step-B is eliminated. In the modification, $|L_{cv}|$ is discarded if no value is removed from $\mathbf{m}_c^{(g-1)}$ and $|L_{cv}|$ is larger than the second value in $\tilde{\mathbf{m}}_c^{(g)}$. This results in the *simplified three-min column-layered decoding*. In this case, a check node requires 3 equal-comparisons to remove the old $|L_{cv}|$ from vector $\mathbf{m}_c^{(g-1)}$ in horizontal Step-A and requires 2 regular comparisons to update $\mathbf{m}_c^{(g)}$ vector with the new $|L_{cv}|$ in horizontal Step-B.

The major difference between the three-min decoding and the simplified three-min decoding is that the three-min decoding requires 3 comparisons for sorting in horizontal Step-B. The simplified three-min requires 2 comparisons for sorting in horizontal Step-B. Our simulation shows that the additional approximation introduced by the simplified three-min decoding only causes very small performance loss. Table I shows that a sorted vector $\mathbf{m}_c$ only gets updated about 4 times in an iteration of the three-min decoding. On the contrary, for TPMP or row-layered decoding, the sorting computation to find the smallest and the second smallest magnitudes for a check node re-starts in every iteration. For the same LDPC code, the average number of updating activities for a sorted vector is more than 16 per iteration. It leads to significant power savings for the three-min column-layered decoding in check node message updating.

TABLE I.    THE AVERAGE NUMBER OF $\mathbf{m}_c^{(g)}$ UPDATING EVENTS IN ONE ITERATION

|  | In the 3rd iteration | In the 6th iteration |
|---|---|---|
| $\widetilde{\mathbf{m}}_c^{(g)} \neq \mathbf{m}_c^{(g-1)}$ | 2.804 | 2.783 |
| $\widetilde{\mathbf{m}}_c^{(g)} = \mathbf{m}_c^{(g-1)}$ , $\lvert L_{cv} \rvert$ being the smallest value in $\mathbf{m}_c^{(g)}$ | 0.036 | 0.050 |
| $\widetilde{\mathbf{m}}_c^{(g)} = \mathbf{m}_c^{(g-1)}$ , $\lvert L_{cv} \rvert$ being the second smallest value in $\mathbf{m}_c^{(g)}$ | 0.140 | 0.170 |
| $\widetilde{\mathbf{m}}_c^{(g)} = \mathbf{m}_c^{(g-1)}$ , $\lvert L_{cv} \rvert$ being the third smallest value in $\mathbf{m}_c^{(g)}$ | 0.880 | 1.005 |
| $\mathbf{m}_c^{(g)} = \widetilde{\mathbf{m}}_c^{(g)} = \mathbf{m}_c^{(g-1)}$ , $\lvert L_{cv} \rvert$ being discarded | 28.140 | 27.992 |

### 2.2.3    Pipelining of Column-layered Decoding

Pipelining is a common practice in VLSI implementation to increase clock speed and thus to speed up data processing throughput. In general, pipelining can only be applied to feed-forward data paths in order to maintain the original function of VLSI circuitry. In LDPC column-layered decoding algorithms, data dependency exists between consecutive layers. Thus, the VLSI circuitry for check node, variable node, and message memories forms a logic loop and pipelining can not be directly applied to increase the effective clock speed of LDPC decoders. In this section, a relaxed pipelining scheme of column-layered LDPC decoding is proposed.

In the original column-layered decoding, the change between $\mathbf{m}_c^{(g-1)}$ and $\mathbf{m}_c^{(g-1+P)}$ is very small if the value of $P$ is not large. Thus, $\mathbf{m}_c^{(g-1)}$ can be used as the estimation of $\mathbf{m}_c^{(g-1+P)}$. An approximation of $R_{cv}^{(g+P)}$ can be calculated from $\mathbf{m}_c^{(g-1)}$ before $\mathbf{m}_c^{(g-1+P)}$ are obtained. Then, it is immediately used for computing $L_{cv}^{(g+P)}$. The approximation allows $P$ clock cycles to complete the message computation for each layer. When $|L_{cv}^{(g+P)}|$ is obtained, the $\mathbf{m}_c^{(g-1+P)}$ is already available. Thus, the horizontal step for layer $g+P$ can be undertaken with $\mathbf{m}_c^{(g-1+P)}$ and $|L_{cv}^{(g+P)}|$. The updating method of $\mathbf{m}_c^{(g+P)}$ is the same as before. The pipelined column-layered decoding algorithm is formulated as the following.

### The Pipelined Column-layered Decoding

---

**Initialization:**

Let $L_{cv} = I_v$ for all variable nodes. For each check node, sort the magnitudes of the $L_{cv}$ messages from its neighboring variable nodes. Compute the sign product for each check node $S_c = \prod_{n \in N(c)} sign(L_{cn})$.

**Iterative decoding:**

For *iter* = 1, 2, …, maximum iteration number

{

  For g=0, 1, …, *G-1*

  {

      ***Compute*** $R_{cv}^{(g+P)}$: For each check node $c$ that connects to variable node $v \in N^{(g)}$, compute $\tilde{\mathbf{m}}_c^{(g+P)}$ by removing the old $|L_{cv}^{(g+P)}|$ from $\mathbf{m}_c^{(g-1)}$. Then, $R_{cv}^{(g+P)} = \tilde{S}_c^{(g+P)} \times m_1$, *where*

$$\tilde{S}_c^{(g+P)} = S_c^{(g-1)} \times old\, sign(L_{cv}^{(g+P)}). \tag{8}$$

      ***Vertical Step***: For each variable node $v \in N^{(g+P)}$, computes $L_{cv}^{(g+P)}$ and $L_v^{(g+P)}$ using (2) and (3).

      ***Horizontal Step-A:*** for each check node $c$ that connects to variable node $v \in N^{(g)}$, compute $\tilde{\mathbf{m}}_c^{(g)}$ by removing the old $|L_{cv}^{(g)}|$ from $\mathbf{m}_c^{(g-1)}$, $\tag{9}$

      ***Horizontal Step-B***: for each check node $c$ that connects to variable node $v \in N^{(g)}$,

compute $\mathbf{m}_c^{(g)}$ using $\tilde{\mathbf{m}}_c^{(g)}$ and $\left| L_{cv}^{(g)} \right|$, $\hspace{3em}$ (10)

and $S_c^{(g)} = \tilde{S}_c^{(g)} \times sign(L_{cv}^{(g)})$. $\hspace{3em}$ (11)

}

**Hard decision and termination:**

Make hard decision by using the sign of $L_v$; Terminate decoding if a valid codeword is found or the maximum decoding iteration is reached.

}

---

### 2.2.4 *Impact to the Complexity of VLSI Implementation*

For the standard Min-Sum based column-layered decoding, a check node requires $d_c - 2$ regular comparators to compute the check-to-variable message for a new column-layer. To execute (1) and (2), the variable-to-check messages corresponding to the whole $\boldsymbol{H}$ matrix must be stored. For the simplified three-min column-layered decoding algorithm, a check node only requires 2 regular comparators. The magnitudes of variable-to-check messages can be used on-the-fly and are not stored. Only the sign bits of variable-to-check messages should be stored. For each check node, three magnitudes, three indices, and one sign bit need to be saved. For a (N, N-M) (4, 32) structured LDPC code, approximately, $(30-2)/30 = 93\%$ of computation can be reduced. Assuming each message is quantized as 4 bits, the number of memory bits needed by a check node is $3 \times (3+5) + 1 = 25$. The percentage of memory bits that can be saved for extrinsic messages is $[(32 \times 4 - 32 - 25) \times M]/(32 \times 4 \times M) = 55.4\%$.

Because the proposed column-layered decoding approaches do not save the magnitude of variable-to-check messages, they are inherently memory-efficient. Equipped with the proposed decoding techniques, an efficient column-layered decoder architecture for quasi-cyclic LDPC codes is developed with architectural and arithmetic optimizations. The details are provided in Section IV.

To evaluate the decoding performance and convergence speed of the proposed algorithms, three LDPC codes are simulated. Codes I and II are rate 1/2, (2304, 1152) LDPC code and rate 5/6, (2304, 1920) LDPC code, respectively, adopted in WiMax (802.16e) standard. Code III is a (4096, 3584) (4, 32)-regular QC-LDPC code, which is constructed using the progressive edge growth method (PEG) [17][18]. The code is also used to illustrate the VLSI architecture design in Section 4. In each simulation, at least 50 frame errors are observed.

Figure 2. shows the frame error rates in decoding the WiMax rate-1/2 (2304, 1152) LDPC code with four different layered decoding algorithms, *i.e.,* 1) the row-layered decoding, 2) the original column-layered decoding, 3) the three-min column-layered decoding, and 4) the simplified three-min column-layered decoding. The maximum number of decoding iterations is set as 10 and 20, respectively, for all decoding approaches. It can be observed that all the four decoding algorithms have almost the same decoding performance with maximum of 20 iterations. With the maximum of 10 decoding iterations, only subtle performance difference can be observed. Figure 3. shows the frame error rates in decoding the WiMax rate-5/6 (2304, 1920) LDPC code with maximum of 10 and 20 iterations, respectively. It shows again that the performance differences among various layered decoding algorithms are negligible.

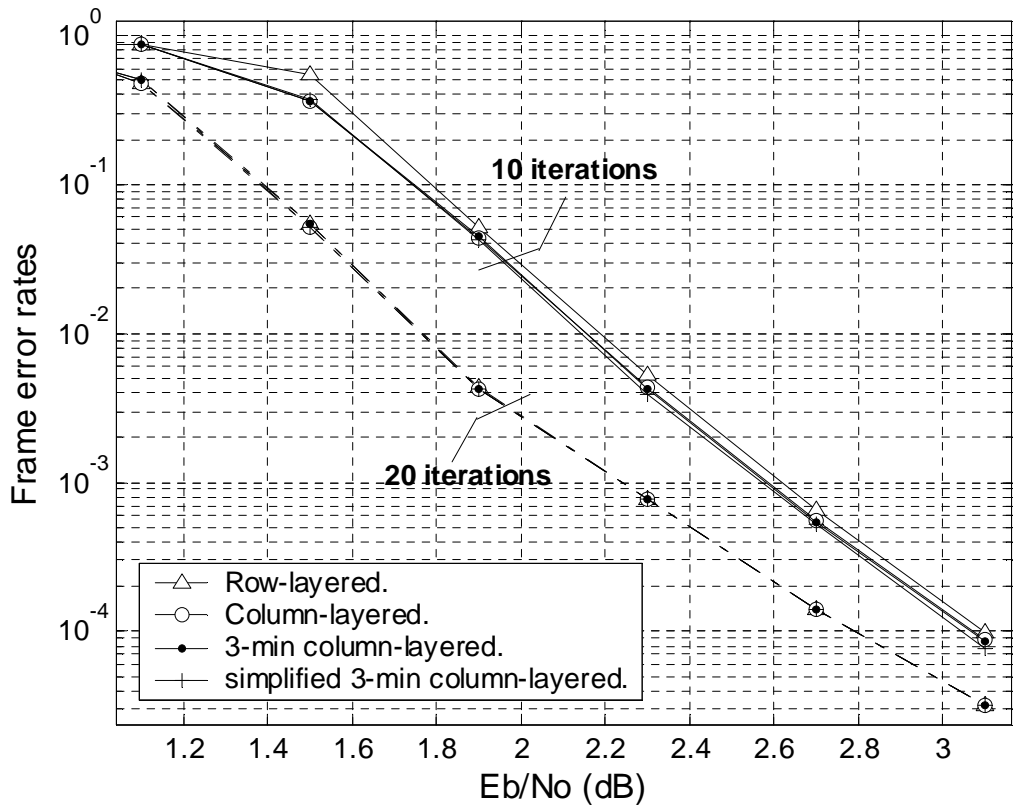

Figure 2.    Frame error rate of various layered decoding approaches in decoding

WiMax rate-1/2 (2304, 1152) LDPC code.

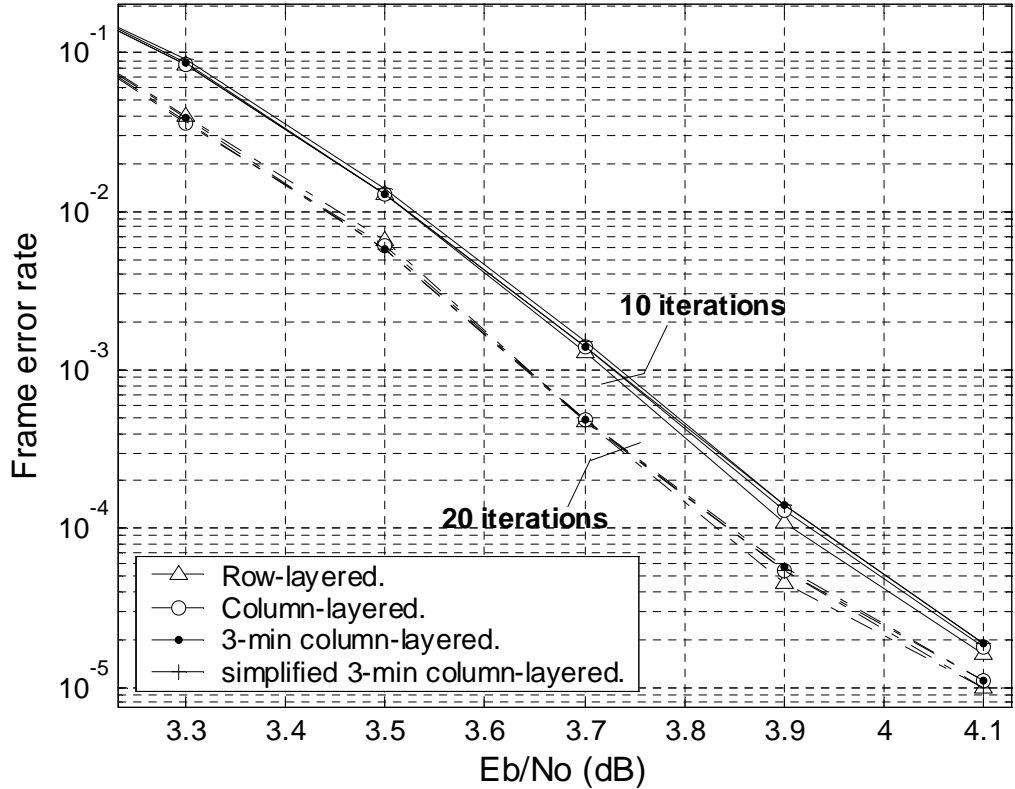

Figure 3.    Frame error rate of various layered decoding approaches in decoding

WiMax rate-5/6 (2304, 1920) LDPC code.

Figure 4. shows the frame error rate (FER) in decoding the (4096, 3584) (4, 32) QC-LDPC code. The number of columns in each column layer is 128. The maximum number of iteration is set as 10. It can be seen that the three-min column-layered decoding algorithm achieves almost the same decoding performance as the standard Min-Sum based column-layered decoding algorithm. The simplified three-min approach has about 0.02dB performance loss. The pipelined three-min column-layered decoding methods introduced slight performance loss compared to the non-pipelined three-min decoding algorithm. With two stages of pipelining, the pipelined decoding scheme has less than 0.02dB performance loss (compared to which??).

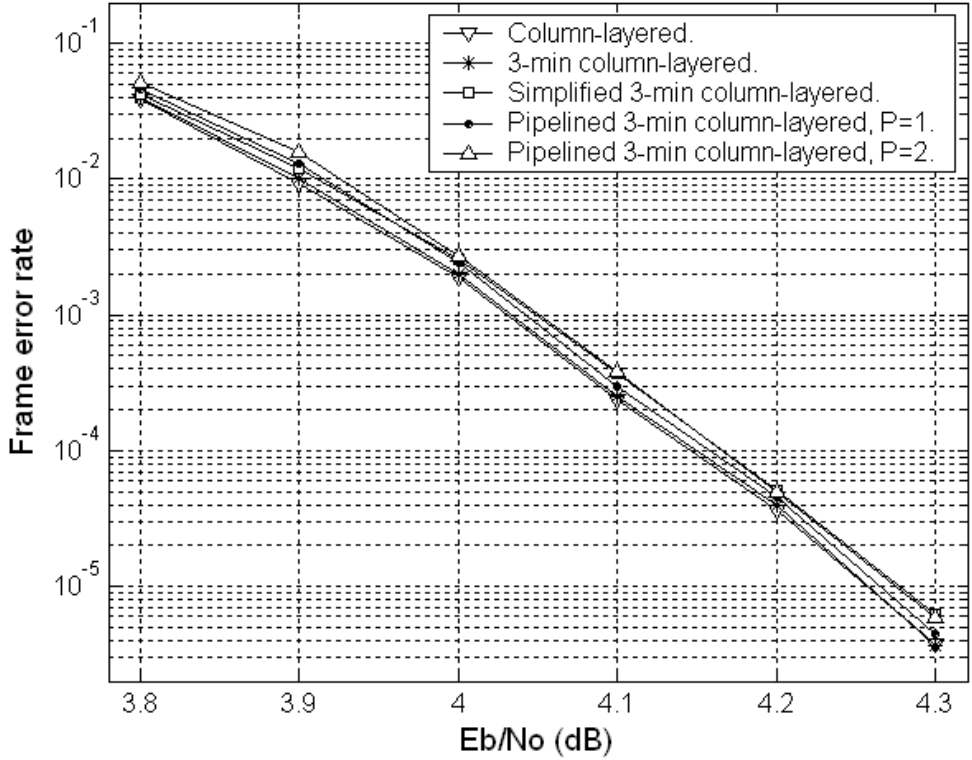

Figure 4.    Frame error rate of various decoding approaches in decoding a (4096, 3584) (4, 32)

QC-LDPC code.

## 4    PROPOSED COLUMN-LAYERED DECODER ARCHITECTURES

In this section, an optimized QC-LDPC decoder architecture for a (4096, 3584) (4, 32)-regular QC-LDPC code using the proposed pipelined column-layered decoding scheme is presented. The architecture efficiently enables very high decoding parallelism for layered decoding while having very short critical path. In one clock cycle, the messages corresponding to 4 circulant matrices are processed in parallel. Two-stage pipelining is employed in order to improve the clock frequency. In order to further reduce the critical path delay, we rearrange the additions in variable node unit to maximally take the advantage of carry-save addition.

*4.1 Column-Layered Decoder Architecture for QC-LDPC Codes*

The parity check matrix of the QC-LDPC code to be considered in this work consists of an array of $4\times32$ cyclically shifted permutation matrices of dimension $128\times128$. The decoding steps discussed in Section II.B are performed for one block column at a time. That means all extrinsic messages corresponding to a block column of the ***H*** matrix are processed in parallel in one clock cycle. The architecture of pipelined decoder with two pipeline stages is shown in Figure 5. Each *M* component in an *M-register* array on the left side of Figure 5. represents a 25-bit register for storing a sorted vector associated with a row of the ***H*** matrix. The *Get Rcv* component performs (8) to get check-to-variable messages for layer *g+P*, where *P=2.* The *CNUa* component (the portion of check node unit for horizontal Step-A) performs (9) to remove the old variable-to-check message associated to layer *g* from the sorted vector. The *CNUb* component (the portion of check node unit for horizontal Step-B) is utilized to perform (10) and (11). The total numbers of *M-register*, *Get Rcv*, *CNUa*, and *CNUb* are all 512. The number of variable node unit (VNU) in the VNU array is 128. A VNU is required to perform the vertical step (2) and (3) for a variable node. The barrel shifters in the left side of the VNU array align the check-to-variable message from row order to column order. Similarly, the barrel shifters in the right side of VNU array align the variable-to-check message from column order to row order. It can be observed form Figure 5. that two types of loops are formed in column-layered decoding. The first type of loop consists of a *CNUa* and a *CNUb* components. The second type of loop is composed of a *Get Rcv*, a *CNUb*, a *VNU*, and two barrel shifters. With the pipelined column-layered decoding approach, the critical path in the second type loop is drastically reduced when applying two-stage pipelining.

The *Get Rcv* component is needed because its outputs are the check-to-variable messages for layer *g+P*. On the contrary, the outputs of *CNUa* are intermediate values for layer *g*. For non-pipelined decoding, the *Get Rcv* component can be eliminated because the output of *CNUa* contains the check-to-variable message for layer *g*. We only need one stage of barrel shifter at the input of CNUa array to minimize critical path. The connections among the array of *CNUa*, *CNUb* and *VNU* have no change during the decoding.

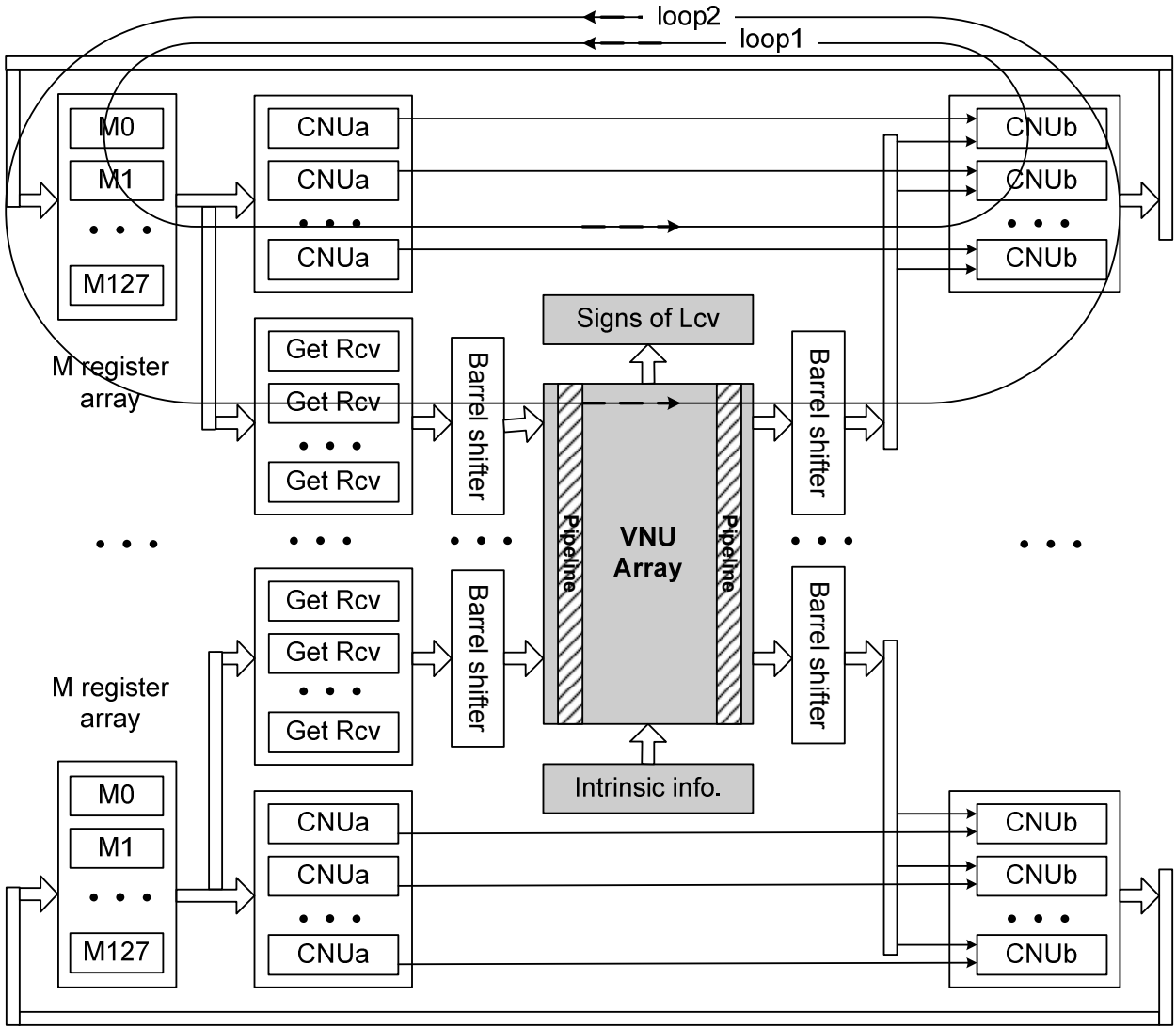

Figure 5.    The top-level block diagram of pipelined column-layered decoding.

## 4.2    Architecture of Check Node Unit

Figure 6. shows the structure of the CNU for the three-min decoding scheme. The sub-matrices in a block row of the **H** matrix are processed one at a time. The CNU is composed of two concatenated stages.

The first stage, *CNUa*, computes the $\tilde{\mathbf{m}}_c^{(g)}$, and the second stage, *CNUb*, generates $\mathbf{m}_c^{(g)}$. The data in Fig. 10 are associated with the vectors in the horizontal decoding step as follows:

$$\mathbf{m}_c^{(g-1)} = [\text{min1\_old}, \quad \text{min2\_old}, \text{min3\_old}],$$

$$\tilde{\mathbf{m}}_c^{(g)} = [\text{min1\_temp}, \quad \text{min2\_temp}, \text{min3\_temp}],$$

$$\mathbf{m}_c^{(g)} = [\text{min1\_new}, \quad \text{min2\_new}, \text{min3\_new}],$$

$$\mathbf{I}_c^{(g-1)} = [\text{idx1\_old}, \quad \text{idx2\_old}, \text{idx3\_old}],$$

$$\tilde{\mathbf{I}}_c^{(g)} = [\text{idx1\_temp}, \quad \text{idx2\_temp}, \text{idx3\_tmp}],$$

$$\mathbf{I}_c^{(g)} = [\text{idx1\_new}, \quad \text{idx2\_new}, \text{idx3\_new}].$$

In each *M-register* for a sorted vector, three smallest magnitudes and their indices are stored. In the decoding of a column layer, if the column index is not in the vector, the magnitudes and indices in the vector are directly passed through the select-logic-A to the second stage. Otherwise, *min1_temp* and *min2_temp* get values from the two remaining smallest magnitudes. The value of *min3_temp* becomes void. The temporary index values are determined in the same way. It is clear that *min1_temp* is the magnitude of the $R_{cv}$ message for the column layer in non-pipelined decoding. After new $L_{cv}$ value is sent back from VNU, it is used to compute the new value for the sorted vector. The structure of a *Get Rcv* component is shown in Figure 6. . Th*is* component is not required for non-pipelined decoding.

For the proposed simplified three-min decoding approach, the adder for the computation of $abs(L_{cv}) - \min 3\_temp$ is not needed. Instead, a simple three-input OR gate can be used to disable *M-register* update in the condition shown in row 5 of Table I. The three inputs for the OR gate are the outputs of the three comparators in *CNUa*. It is shown in Fig. 3 that the removal of the adder results in less than 0.02dB performance loss.

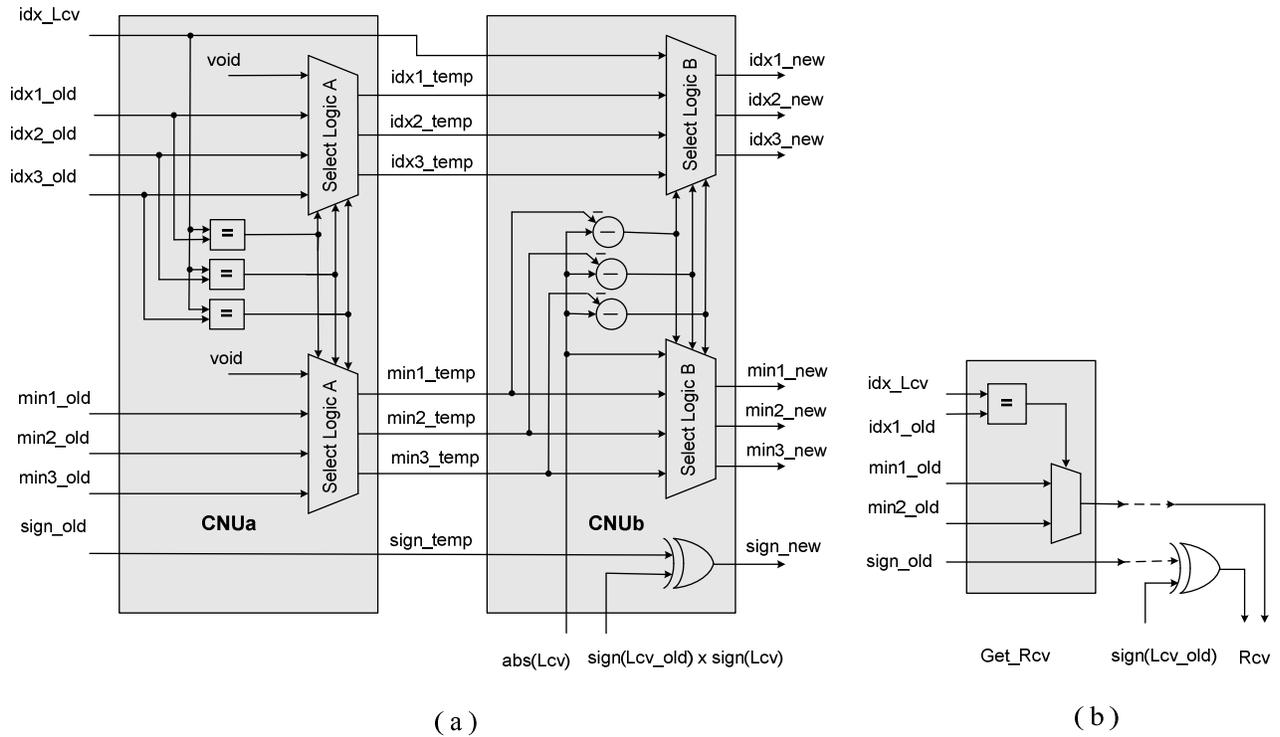

( a )                                                ( b )

Figure 6.        (a) CNU architecture for the three-min column-layered decoding.  (b) The structure of the
*Get Rcv* component.

### 4.3    Architecture of Optimized Variable Node Unit

Shown in Figure 7.  is the structure of an optimized VNU that can simultaneously process 4 check-to-variable messages. The addition operations are rearranged such that the advantage of carry-save adder can be maximally taken. In the beginning of a VNU, the check-to-variable messages in signed-magnitude format are converted to 2's complement representation. In the data conversion for each signed-magnitude number, the sign bit is not immediately added to the bitwise-not of lower bits. Instead, all sign bits are added through adder array. Then, each two-bit sign-sum is sent to an adder in the second addition stage for final summation. Because each adder in the second and third stages has three inputs, it can be implemented using a carry-save adder and a regular binary adder. The right shift operations $>>1$ and $>>2$ are used for performing the scalar multiplication of 0.75. The above mentioned arithmetic optimizations aim to significantly reduce the logic delay in a VNU.

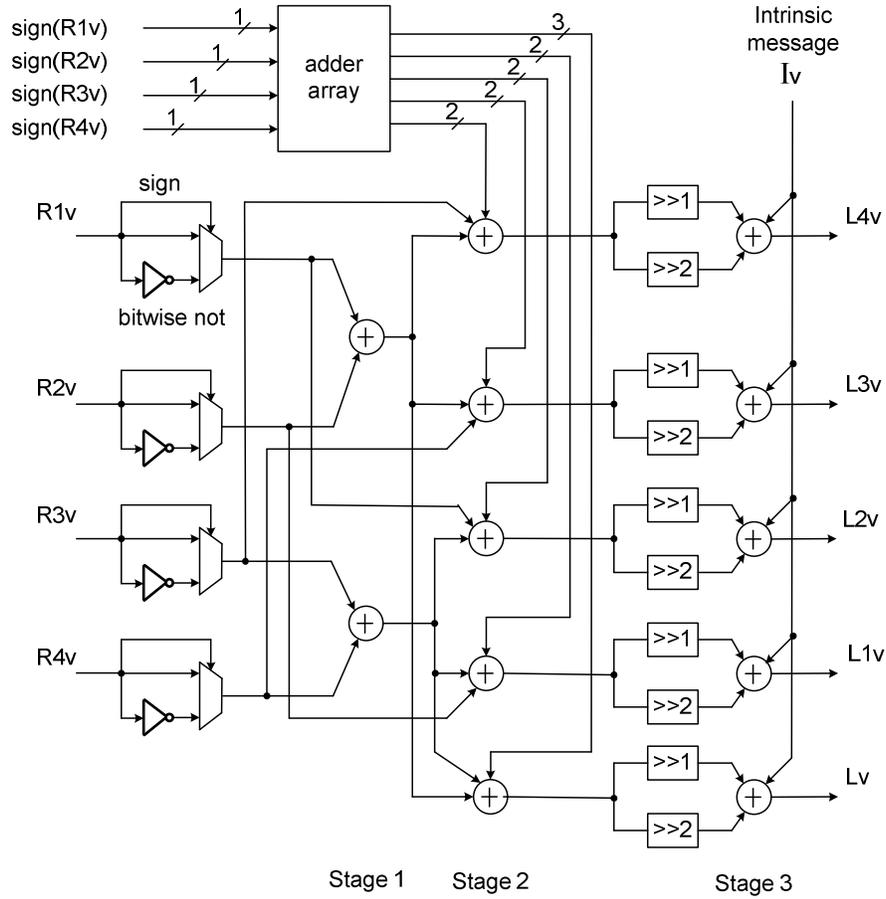

Figure 7.　　The architecture of the optimized VNU .

To illustrate the data flow of the optimized VNU, let us take the computation of $L_{3v}$ as an example. Assume that $R_{4v}$ is a negative number and other inputs are positive numbers. The value before the shift of scaling operation of $L_{3v}$ is $(|R_{1v}|+|R_{2v}|+bit\_inverse(|R_{4v}|)+'0'+'0'+'1')$ . The $'0'+'0'+'1'$ computation is performed by a 1-bit full adder in the adder array block associated with $L_{3v}$ message. After the final shift and addition stage, the output of $L_{3v}$ is $0.75\times(R_{1v}+R_{2v}+R_{4v})+I_{v}$ .

## 5 HARDWARE REQUIREMENT AND DECODING THROUGHPUT

The pipelined column-layered decoder with two pipeline stages is modeled using Verilog RTL and synthesized using Fujitsu 0.13um standard library. The required hardware resource and the synthesis result are summarized in Table II. Each intrinsic message is quantized as 4 bits. It takes 32 clock cycles to compute the initial sorted vectors and $32 \times 10 = 320$ clock cycles for 10 decoding iterations. The number of clock cycles for pipeline latency is 2. Let $f_{clk}$ denote the clock frequency of the decoder, the information (source data) decoding throughput is $f_{clk} \times (4096 - 512)/(32 + 320 + 2)$. Thus, the decoder achieves a decoding throughput of 3.928 Gb/s at a clock speed of 388MHz.

TABLE II.    THE HARDWARE RESOURCE FOR (4096, 3584) (4, 32) QC-LDPC DECODERS

| VNU | 128 |
|---|---|
| CNU | 512 |
| Get_Rcv | 512 |
| Intrinsic message (bits) | $4 \times 4096 = 16384$ |
| Sorted vector (bits) | $25 \times 512 = 12800$ |
| Signs of variable-to-check messages | $4 \times 4096 = 16384$ |
|  |  |
| Area per VNU ( $um^2$ ) | 5597.8 |
| Area per CNU ( $um^2$ ) | 2100.9 |
| Area per Get_Rcv ( $um^2$ ) | 152.1 |
| Clock Frequency (MHz) | 388 |
| Synthesis area ( $mm^2$ ) | 6.755 |
| Information decoding throughput (Gb/s) | 3.928 |

Table III compares the proposed column-layered decoder with the state-of-the-art row-layered decoders. To mitigate the discrepancy introduced from different implementation technologies, the area and clock speed of all these designs are scaled to 65nm. For the implementation technology of 0.18um, 0.13um, and 90nm, the area scaling down factor is set as 8, 4, and 2, respectively. The corresponding clock frequency is scaled up by $1.25^3$, $1.25^2$, and 1.25, respectively. The maximum decoding iteration is set as 10 for all decoders with layered decoding algorithms. For area of synthesis result, a scaling factor of 1/0.7 is applied to approximate the layout area.

| | This work | [11] | [14] | [12] | [13] |
|---|---|---|---|---|---|
| Code | (4096, 3584) | (9600, 7200) | 802.11n | 802.16e, 802.11n | 2048-bits |
| Rate | 7/8 | 3/4 | 1/2 ~5/6 | 1/2 ~ 5/6 | 1/2 ~ 7/8 |
| Algorithm | Column-layered Min-Sum | Row-layered Min-Sum | Row-layered Min-Sum | Row-layered BP | Row-layered BCJR |
| LLR message quantization | 4-bits | 6-bits | 5-bit | - | 4-bits |
| Max. iteration | 10 | 10 | 5 | 10 | 10 |
| Max decoding parallelism | 512 | 80 | 81 | 2×96 | - |
| | | | | | |
| Technology | 0.13um | 65nm | 0.18um | 90nm | 0.18um |
| Clock frequency(MHz) | 388 | 500 | 208 | 450 | 125 |
| Area | 6.755 (synthesis) | 0.504 (synthesis) | 3.39 | 3.5 | 14.3 |
| Max info. throughput (Gb/s) | 3.928 | 1.08 | 0.78 | 1.0 | 0.64 |
| | | | | | |
| Clock frequency (scaled to 65nm) | 606 | 500 | 406 | 562 | 244 |
| Layout area (scaled to 65nm) | 2.41 | 0.72 | 0.42 | 1.75 | 1.79 |
| Normalized info. throughput (scaled to 65nm) | 6.13 | 1.08 | 0.76 (10 iterations) | 1.25 | 1.25 |

Considering the normalized decoding throughput, throughput/area ratio, and decoding performance among various designs, it can be concluded that the proposed simplified column-layered decoding algorithm and architecture have significant advantages in high throughput LDPC decoder implementation.

## 6   CONCLUSION

In this paper, various techniques have been explored to reduce the computation complexity of the column-layered decoding. As a result, the proposed method can drastically reduce the overall computation complexity of the original scheme while largely maintaining decoding performance and convergence speed. In addition, a relaxed pipelining scheme has been shown to break the data dependency between adjacent column layers, and thus enhance the clock speed. Combining all the proposed techniques, a low-complexity, high-speed LDPC decoder architecture for generic QC-LDPC codes has been developed and the implementation result for a specific example has demonstrated that the proposed column-layered decoder architecture is very competitive to state-of-the-art row-layered LDPC decoder designs.